\documentclass[twocolumn,preprintnumbers,amsmath,amssymb,floatfix,superscriptaddress,prb,showpacs]{revtex4}

\usepackage{graphicx}
\usepackage{dcolumn}
\usepackage{bm}

\begin{document}

\title{Nonlinear Control of Tunneling Through an Epsilon-Near-Zero Channel}

\author{David A. Powell}
\email{david.a.powell@anu.edu.au}
\affiliation{Nonlinear Physics Center, Research School of Physics and Engineering, Australian National University, Canberra ACT 0200, Australia}
\author{Andrea Al\`u}
\affiliation{Department of Electrical and Systems Engineering, University of Pennsylvania, Philadelphia, Pennsylvania 19104, USA}
\affiliation{The University of Texas at Austin, Department of Electrical \& Computer Engineering, Austin, TX 78712-0240, USA}
\author{Brian Edwards}
\author{Ashkan Vakil}
\affiliation{Department of Electrical and Systems Engineering, University of Pennsylvania, Philadelphia, Pennsylvania 19104, USA}
\author{Yuri S. Kivshar}
\affiliation{Nonlinear Physics Center, Research School of Physics and Engineering, Australian National University, Canberra ACT 0200, Australia}
\author{Nader Engheta}
\affiliation{Department of Electrical and Systems Engineering, University of Pennsylvania, Philadelphia, Pennsylvania 19104, USA}

\begin{abstract}
The epsilon-near-zero (ENZ) tunneling phenomenon allows full transmission of waves through a narrow channel even in the presence of a strong geometric mismatch. Here we experimentally demonstrate nonlinear control of the ENZ tunneling by an external field, as well as self-modulation of the transmission resonance due to the incident wave.  Using a waveguide section near cut-off frequency as the ENZ system, we introduce a diode with tunable and nonlinear capacitance to demonstrate both of these effects.  Our results confirm earlier theoretical ideas on using an ENZ channel for dielectric sensing, and their potential applications for tunable slow-light structures.
\end{abstract}


\maketitle

\section{Introduction}

One of the most active topics in current electromagnetics research is the study of artificial structures where permittivity $\epsilon$ and permeability $\mu$ are engineered.  This allows them to have values not available in natural materials, or which cannot normally be obtained at the desired operating frequency.  A particularly interesting example is materials having a permittivity of zero at some frequency \cite{Alu2007a,Lovat2006,Lovat2007}.  In such structures the wavelength becomes infinite, and wave propagation over distances much larger than the free space wavelength can be treated as quasi-static.  Since this qualitative behavior is still observable even when the complex permittivity is close to but not identically zero, the term epsilon-near-zero (ENZ) has been coined to describe it.

As one interesting application of ENZ materials, it has been shown theoretically that a narrow ENZ channel would support complete transmission of a signal incident from a larger waveguide \cite{Silveirinha2006}, despite the large geometric mismatch.  This was subsequently demonstrated experimentally in systems where the ENZ response was engineered via a surface pattern \cite{Liu2008}, and also using the natural dispersion characteristics of a rectangular waveguide near its cut-off frequency without the use of any composite structure \cite{Edwards2008a}.

Using the cut-off waveguide approach, it is straightforward to tailor the center-frequency and bandwidth of this transmission effect by modifying the geometry of the waveguide.  It has also been shown that the ENZ tunneling frequency is sensitive to a dielectric cavity included within the waveguide, thus it can be used for sensing~\cite{Alu2008}.  This paper aims to demonstrate experimentally that the tunneling effect can be dynamically controlled by placing a tuning element within the waveguide.  This allows for efficient control of the ENZ transmission, with potential applications in tunable slow-light structures.  We also demonstrate that the introduction of nonlinearity into the system allows the ENZ resonance to be controlled by the incident wave itself.

\section{Tuning the ENZ transmission}

\begin{figure}[htb]
\includegraphics[width=\columnwidth]{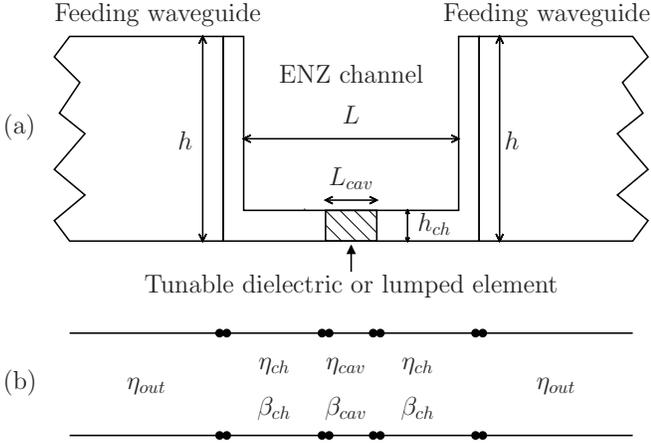}
\caption{(a) Structure of the ENZ channel, (b) equivalent circuit model. }
\label{fig:structure}
\end{figure}

The structure under consideration is presented in Fig.~\ref{fig:structure}(a), where rectangular waveguides of width $w$ and height $h$ feed a signal through a narrow section of width $w$, height $h_{ch}$, and length $L$, which exhibits the ENZ property at microwave frequencies.  Since the ENZ tunneling occurs at the cut-off frequency of the fundamental TE$_{10}$ mode of the narrow section, the feeding sections of waveguide are made from a material with a higher dielectric constant to ensure that their TE$_{10}$ mode is propagating.  In Refs.~\onlinecite{Alu2008,Alu2008d} an equivalent circuit model was given for the structure, with the feeding waveguides, ENZ channel and dielectric cavity represented by sections of transmission line, as shown in Fig.~\ref{fig:structure}(b). The reactances at the transition due to the excitation of evanescent higher-order modes \cite{Marcuvitz1951} have previously been shown to have a negligible effect on the ENZ tunneling.  Equation~\eqref{eq:sens_dist} was then derived as the condition for full transmission of an ENZ system including a cavity with a different dielectric constant from the rest of the ENZ channel:
\begin{equation} \label{eq:sens_dist}
 \frac{\eta_{out}^{2}}{s^{2}\eta_{ch}^{2}} = 1-\frac{2\Delta\sin \theta}
{\Delta + \Delta\cos \varphi \sin \theta +2\eta_{cav}\eta_{ch}\cos \theta \sin \varphi}.
\end{equation}
Here $\Delta = \eta_{ch}^{2}-\eta_{cav}^{2}$, $\theta = \beta_{cav}L_{cav}$, $\varphi = \beta_{ch}\left(L-L_{cav}\right)$, $\eta$ and $\beta$ represent the impedance and wavenumber of each section and $s = h_{ch}/h$.  The subscript $ch$ refers to the ENZ channel, and $cav$ to the dielectric cavity within it, and all relevant dimensions are given in Fig.~\ref{fig:structure}(a).  Equation~\eqref{eq:sens_dist} has one solution associated with ENZ tunneling, and an infinite number of solutions corresponding to the Fabry-Perot resonances along the channel. It can be used to show how the ENZ transmission frequency will change due to a change in dielectric constant of the cavity, which could potentially include the whole length of the ENZ section.  This could be used in a sensing configuration where the cavity is filled with some material of unknown dielectric constant, or alternatively for tuning if the dielectric constant can be controlled externally.

\begin{figure}[htb]
\includegraphics[width=\columnwidth]{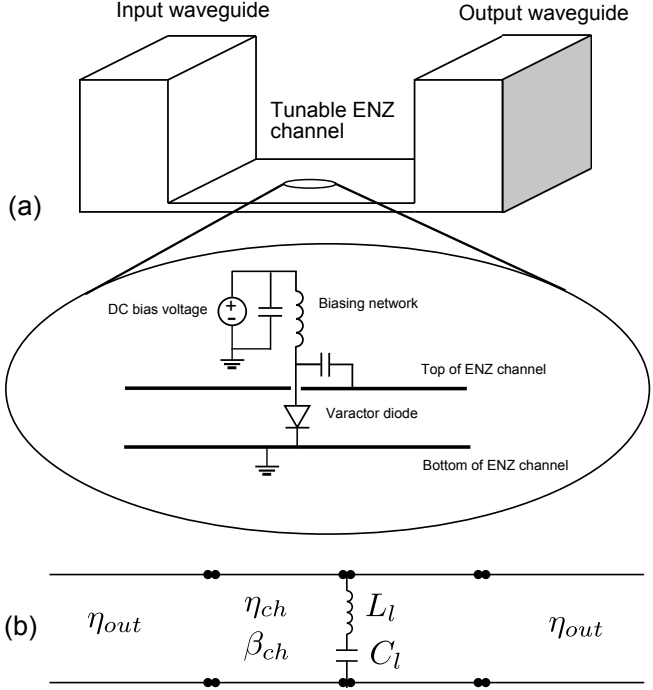}
\caption{(a) The ENZ channel with a lumped tunable inclusion, and (b) the corresponding equivalent circuit model.}
\label{fig:lumped_structure}
\end{figure}

At microwave frequencies, tuning is much more readily achieved using lumped nonlinear components, as illustrated in Fig.~\ref{fig:lumped_structure}(a).  This can be regarded as a limiting case of a dielectric inclusion of small cross-section, however it is much more convenient to include the lumped elements into the equivalent circuit model, as shown in Fig.~\ref{fig:lumped_structure}(b)

\begin{figure}[htb]
\includegraphics[width=\columnwidth]{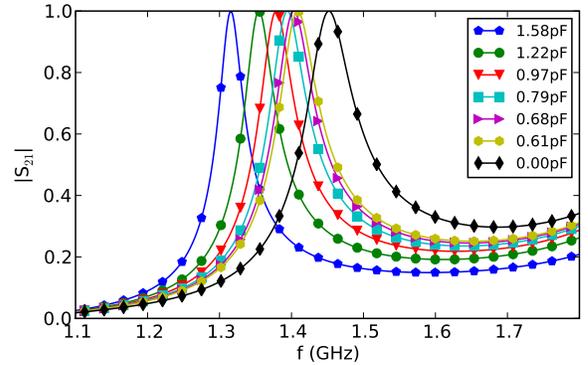}
\caption{The equivalent-circuit transmission response of an ENZ channel loaded with a lumped capacitance.\label{fig:tuning_ideal}}
\end{figure}

We are most interested in capacitive lumped inclusions, however the insertion of a narrow conducting element between the top and bottom of the waveguide will also result in some additional inductance in the equivalent circuit model \cite{Marcuvitz1951}, due to the excitation of higher-order evanescent modes.  Thus we consider here a series inductance $L_{l}$ and capacitance $C_{l}$ with a shunt connection between the transmission line section, noting that the values in the equivalent circuit model also include the effect of parasitics and the excitation of higher-order modes.  The resultant expression for complete transmission is:
\begin{equation}\label{eq:sens_lump}
\frac{\eta_{out}^2}{s^2\eta_{ch}^2} = 1 +\frac{2\omega\eta_{ch}C_{l}}
{2\left(1-\omega^{2}L_{l}C_{l}\right)\sin L\beta_{ch}+\omega\eta_{ch}C_{l}\left(1+\cos L\beta_{ch}\right)}.
\end{equation}
This also has solutions for the ENZ and Fabry-Perot resonances, and generalizes the dielectric sensing and tunability of this structure to include lumped elements.  Thus the study of lumped nonlinear inclusions can yield much insight into the broader behavior of nonlinear ENZ systems.  It should be noted that the impedance $\eta$ of a waveguide mode is not uniquely defined \cite{Marcuvitz1951}, however for consistency with the circuit model of Fig.~\ref{fig:lumped_structure} we are required to use the following definition:
\begin{equation}\label{eq:impedance}
\eta = 2 h\omega\mu_{0}/w\beta
\end{equation}
We have confirmed numerically that the introduction of a lumped linear capacitance can be used to tune the frequency of the ENZ response.  Figure~\ref{fig:tuning_ideal} shows the transmission response with $L_{l}=3.5$nH as $C_{l}$ is varied, calculated using our equivalent circuit model.

It can be seen that the maximum magnitude of the transmission is not changed, but the resonant frequencies are shifted and the quality factor is increased by the additional capacitance.  The electric field within the channel at the ENZ transmission frequency was analyzed in the commercial electromagnetics solver CST Microwave Studio \cite{CST} for a channel loaded with a 4pF capacitor.  It was found that a phase variation of only 4$^{\circ}$ occurred over the length of the channel, which is one third of the free space wavelength.  The complete transmission and low phase variation confirm that the essential features of the ENZ coupling have been maintained.

\section{Experimental demonstration of tunability}

To confirm the tunability, we use an experimental configuration and parameter retrieval procedure similar to that described in Ref.~\onlinecite{Edwards2008a}, with $L=99$mm, $h=50.8$mm and $w=101.6$mm.  The input and output waveguides consist of PTFE (Teflon) beams covered in conductive tape, fed by a coaxial connector attached to a probe.  Smaller probes are placed within these waveguides to measure the amplitudes of the modes propagating inside the waveguide and to find the transmission and reflection parameters of the ENZ channel.  The transmission through the system is measured at 1601 frequency points using a network analyzer, and by measuring the amplitude and phase of all the incident and reflected waves within the feeding waveguides, we are able to solve for the scattering parameters of the ENZ channel.

The ENZ channel itself is fabricated from brass sheet and blocks.  To introduce a tunable capacitance, we use an SMV1231 varactor diode, with capacitance tunable between 0.45-2.35pF.  This is soldered to a small piece of circuit-board introduced into the ENZ channel.  As shown in Fig.~\ref{fig:lumped_structure}(a) there is a direct electrical connection at one end and a small inductor-capacitor network at the other to allow DC biasing whilst still maintaining a low impedance RF connection to the channel wall.

\begin{figure}[htb]
\includegraphics[width=\columnwidth]{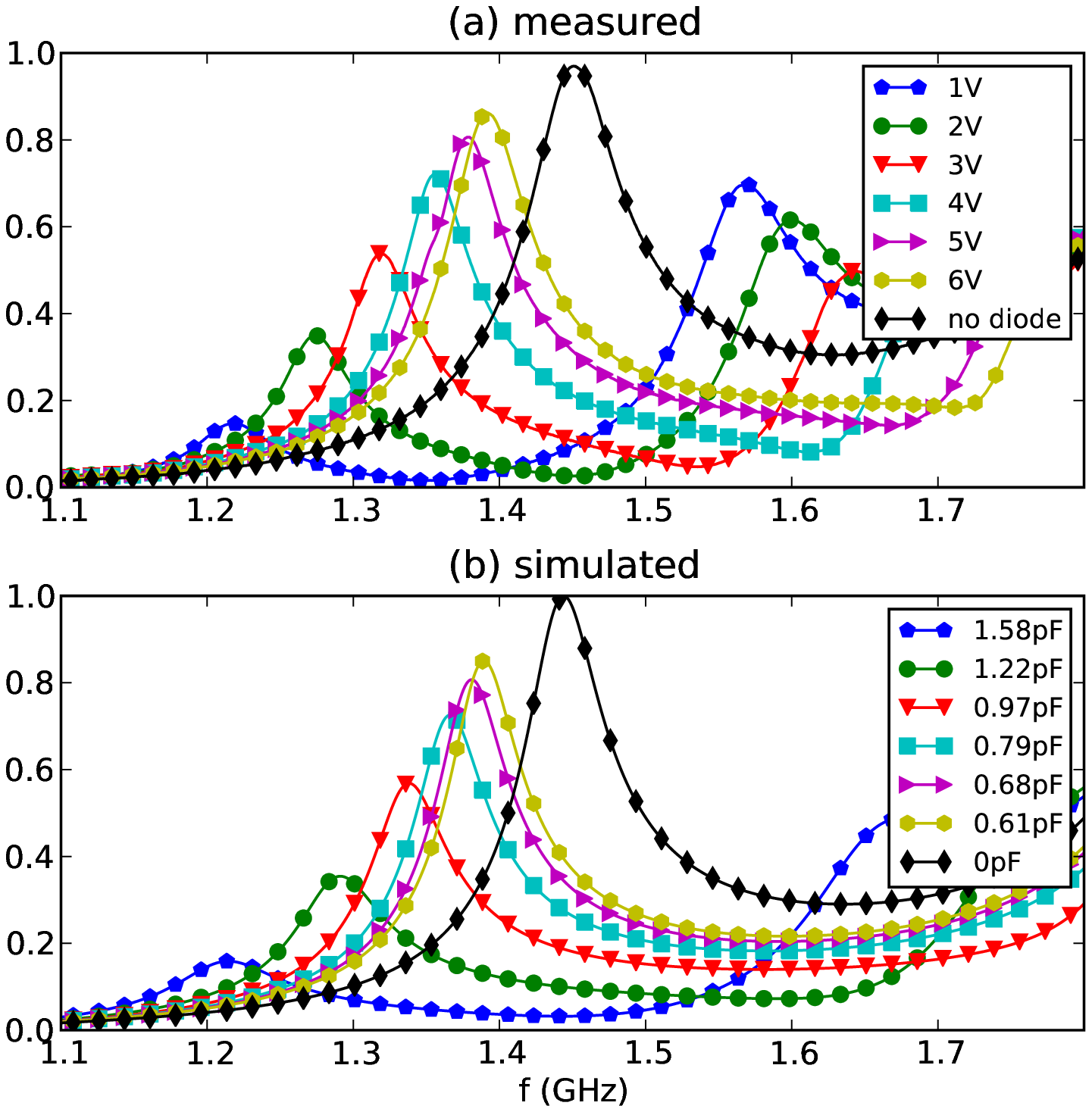}
\caption{Tuning of ENZ transmission with varactor diode for channel height of 3.18mm: (a) experimental, (b) numerical results with corresponding capacitance.\label{fig:tuning_narrow}}
\end{figure}

Figure~\ref{fig:tuning_narrow}(a) shows the measured transmission as a result of tuning the DC voltage and hence the diode capacitance.  The observed frequency shift is similar to that predicted by the equivalent circuit model, as shown in Fig.~\ref{fig:tuning_ideal}, although there is as a reduction in the transmission amplitude and quality factor for higher capacitance values.  The discrepancy is due to the parasitic capacitance, inductance and resistance of the diode, the inductance due to the interruption in the waveguide, as well as the influence of the circuit board and biasing components.  However, it is clearly shown that high transmission can still be achieved for a substantial tuning range, and that for quite reasonable values of voltage the ENZ resonance can effectively be switched on and off by the control signal.

The numerical results obtained from CST Microwave Studio are shown in Fig.~\ref{fig:tuning_narrow}(b) for comparison, corresponding to the diode capacitance for each voltage in Fig.~\ref{fig:tuning_narrow}(a).  By adding a parasitic inductance of 3.5nH and a parasitic capacitance of 0.1pF due to the circuit board and biasing circuit, we are able to achieve reasonable agreement with experiments.  We note that exact agreement is difficult due to the high sensitivity of the channel resonance to small variations in component values and to minor geometrical imperfections.

\begin{figure}[htb]
\includegraphics[width=\columnwidth]{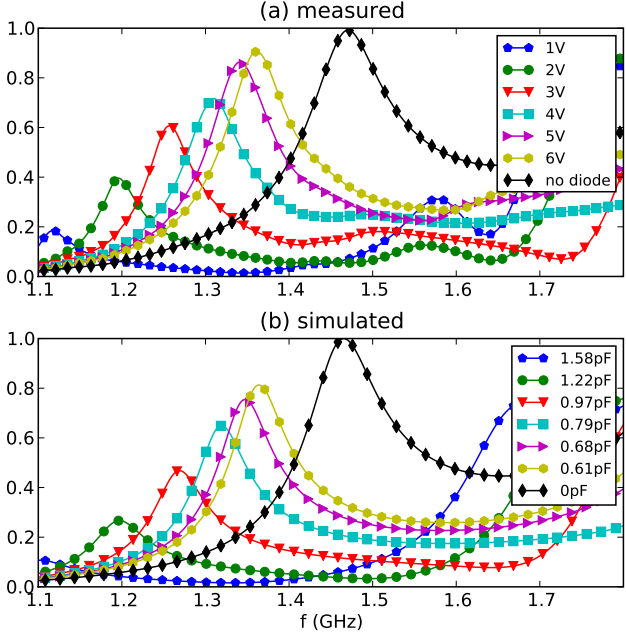}
\caption{Tuning of ENZ transmission with varactor diode for channel height of 5.56mm: (a) experimental, (b) numerical results with corresponding capacitance.\label{fig:tuning_wide}}
\end{figure}

The equivalent circuit results presented in Ref.~\onlinecite{Alu2008} show that the sensitivity of the channel to the dielectric change is independent of the height of the ENZ waveguide section \footnote{This can be understood from Eq.~\eqref{eq:sens_dist} by noting that the impedance $\eta$ defined in Eq.~\eqref{eq:impedance} scales as $1/h$.  We see that the right hand side of the equation remains constant, since $s$ and $\eta_{ch}$ scale proportionally and inversely with $h_{ch}$ respectively.  In the fraction on the right hand side, the terms $\Delta$ and $\eta_{ch}\eta_{cav}$ both scale as $1/h_{ch}^2$, thus it also remains unchanged.  In practice there would be a small frequency change due to the shunt susceptance at the discontinuities which we neglect here.}.  In contrast, the sensitivity to a lumped capacitance is strongly dependent on the height of the waveguide, since this modifies $\eta_{ch}$, but has only a small effect on $L_{l}$ and $C_{l}$ due to evanescent higher-order modes.  To demonstrate this effect we have repeated the experimental and numerical results with a another channel, as shown in Fig.~\ref{fig:tuning_wide}.  It can be seen that the increase in the channel height improves the tunability of the structure.  To achieve agreement with experimental results, the parasitic inductance was changed to 2.6nH, to reflect the necessarily different dimensions of the biasing structure in this experimental configuration.

\begin{figure}[htb]
\includegraphics[width=\columnwidth]{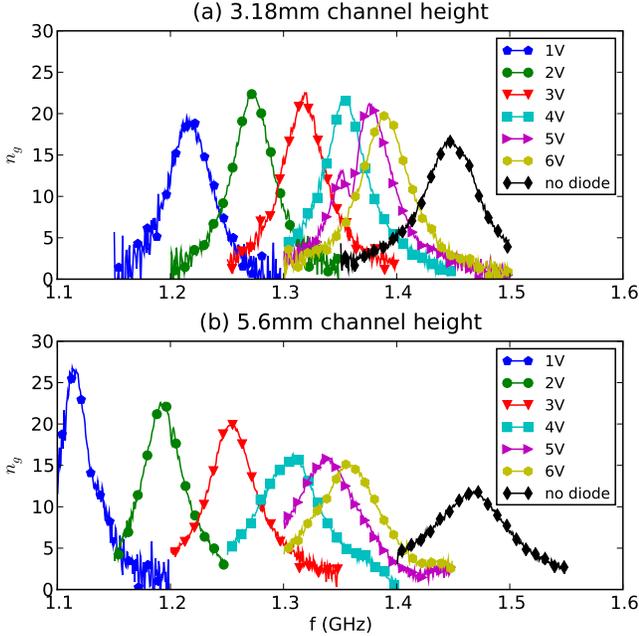}
\caption{Experimental tuning of group index for channel height of (a) 3.18mm and (b) 5.56mm.\label{fig:tuning_delay}}
\end{figure}

\section{Group velocity}

As this structure is strongly resonant and exhibits a high phase velocity, it is clear that it should have a correspondingly low group velocity.  Since the ENZ tunneling theory is also applicable in plasmonic systems \cite{Alu2008c}, a nonlinear ENZ channel coupled to external waveguides could be used as a tunable slow-light structure.  In order to demonstrate this effect and to analyze the response of this system to a signal with finite bandwidth, we have calculated the group delay from the experimentally measured transmission.  The effective group index, calculated as $n_{g}(\omega) = cL^{-1}\frac{d}{d\omega}\arg\left(S_{2,1}\right)$, is shown in Fig.~\ref{fig:tuning_delay}.  It should be noted that this quantity is calculated for the finite structure rather than an infinite ENZ channel, as discussed in Ref.~\onlinecite{Silveirinha2006}.

The effective group index can reach 25 for the measured structures, and the maximum delay corresponds to the transmission peak as expected.  To put these results in context, we consider the frequency variation of the imaginary part of the input impedance of the system, which closely corresponds to the group delay.  At the frequency of ENZ tunneling with no tuning elements, it can be evaluated in closed form as:
\begin{equation}
\frac{d}{df}\mathrm{Im}(Z_{in}) = \frac{2\mu_{0}h_{ch}}{(w \pi k_{e})^3}\left(s^2-1\right)\sin\left(2 k_{e} L\right),
\end{equation}
where the effective wavenumber $k_{e}$ is given by:
\begin{equation}
k_{e} = \pi \frac{h_{ch}}{w}\sqrt{\frac{\epsilon_{out}-\epsilon_{ch}}{h_{ch}^2\epsilon_{out}-h^2\epsilon_{ch}}}
\end{equation}
and $\epsilon_{out}$ and $\epsilon_{ch}$ are the dielectric constants of the media filling the output and ENZ channels, respectively.  This suggests that to first order reducing the height $h_{ch}$ of the ENZ section will increase the delay proportionally.  This is confirmed by the results in Fig.~\ref{fig:tuning_delay} for the structures without a diode, however the additional losses introduced in the tuned structures complicate the picture somewhat.

\begin{figure}[htb]
\includegraphics[width=\columnwidth]{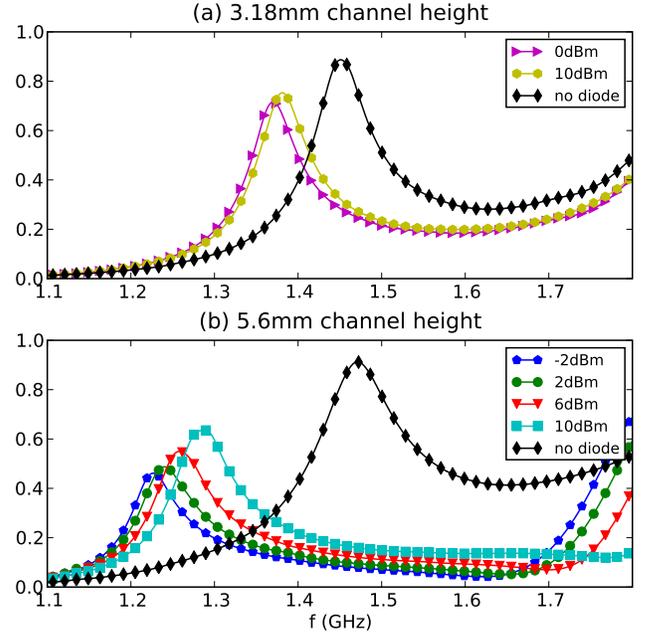}
\caption{Experimental self-modulation of ENZ transmission by the input wave for diode-loaded channels of height of (a) 3.18mm and (b) 5.56mm \label{fig:self_tuning}}
\end{figure}

\section{Nonlinear self-Modulation of the ENZ transmission}

External control of ENZ tunneling has a great deal of potential to increase the flexibility and utility of the phenomenon.  An even more exciting prospect is that the impinging field can self-modulate the ENZ response.  A narrow channel containing a nonlinear dielectric material will show an enhanced nonlinear response due to the strong field confinement.  The phase uniformity of the transmission offers interesting opportunities for multi-frequency nonlinear processes such as harmonic generation and parametric amplification.  Since phase matching conditions are highly critical to the efficiency of these processes, the essentially quasi-static nature of the fields at the ENZ frequency should be of great benefit for the required dispersion engineering.  Here we show that a nonlinear inclusion made from a pair of oppositely oriented diodes in series exhibits a self-modulation property, and thus serves as a proof-of-concept for nonlinear ENZ tunneling.  The principle is equally applicable if the material parameters are based on those of resonant metamaterial composites, with many authors showing that their electric or magnetic resonant frequency can be controlled by an external or incident field (e.g. Refs.~\onlinecite{Gorkunov2004,Gil2004,Powell2007}).  Figure~\ref{fig:self_tuning} shows the results of this self-tuning effect.

Increasing the incident power has the effect of reducing the effective capacitance of the diode and hence bringing transmission closer to the situation without the diode, consistent with the application of a higher DC bias.  It is also clear that we achieve a lower level of tuning in the narrow channel.  Again this is consistent with the DC tuning results, however it is difficult to quantify this effect.  This is due to our feed structure which does not provide uniform coupling to the waveguide over the whole frequency band, hence the incident power on the ENZ section has strong variation with frequency.

\section{Conclusion}

We have demonstrated experimentally that the epsilon-near-zero tunneling effect can be dynamically controlled by introducing a lumped nonlinear element.  This allows the frequency of ENZ tunneling to be tuned or suppressed by an external signal, which generalizes previous results showing the sensitivity of the phenomenon to a dielectric change within the cavity.  In addition, the nonlinearity of the inclusion allows the response to be modulated by an impinging wave, which combined with the low phase delay across the structure should allow the observation of many interesting nonlinear effects. 

\section*{Acknowledgments}

DP thanks Prof N. Engheta and his group for their warm hospitality during his visit to the University of Pennsylvania.
DP and YK acknowledge support from the Australian National University and the Australian Research Council.


\end{document}